\documentstyle[sprocl,epsfig]{article}

\def\Journal#1#2#3#4{{#1} {\bf #2}, #3 (#4)}

\def\apj{\em ApJ}
\def\apjl{\em ApJL}
\def\aa{\em A\&A}
\def\mnras{\em MNRAS}

\def\be{\begin{equation}}
\def\ee{\end{equation}}
\def\bea{\begin{eqnarray}}
\def\eea{\end{eqnarray}}

\begin{document}

\title{MODES OF DISC ACCRETION ONTO BLACK HOLES}

\author{A. MERLONI}

\address{
Max-Planck-Insitut f\"ur Astrophysik,
  Karl-Schwarzschild Str. 1, D-85741, Garching, Germany}


\maketitle\abstracts{I review our current theoretical understanding
  of the different possible rotating modes of accretion onto a black hole. I
  discuss both thick adiabatic flows and radiatively efficient thin disc
  solutions. I present a new self-consistent unified analytical description of two-phase
  thin disc--corona systems in which magneto-rotational instability is responsible
  for angular momentum transport in the disc and for the corona generation. 
Finally, I briefly discuss the role of magnetic fields in bridging the
  gap between accretion discs theory and jet production mechanisms.}

\section{Accretion discs around black holes: basics}
\label{sec_acc_theo} 
\noindent

Accretion is the physical process by which celestial bodies aggregate matter
from their surroundings. In the case of compact objects, 
the gravitational binding energy that such matter
must release for accretion to occur is a powerful source of luminosity.
If the accreting material possesses angular momentum, though, gravitational contraction
is impeded by a centrifugal barrier, as specific angular momentum is 
hard to dispose of, and the formation of a disc is unavoidable. It is
such a
disc that mediates the process of accretion: neighboring annuli of differentially 
rotating matter experience a viscous shear (of some kind) that transports angular momentum
outwards and allows matter to slowly spiral in towards the center of the
potential. 

The variety of spectral energy distributions  
from accretion-powered objects is a clear sign that
different accretion modes are possible. A fundamental discriminant is
the radiative efficiency of the flow, defined as the  ratio of the
output bolometric luminosity 
to the rest mass energy accretion rate, $\epsilon \equiv
L/\dot M c^2$.
When radiative cooling is
efficient (standard `Shakura-Sunyaev' solution, hereafter
SS \cite{ss73}), the maximal energy per unit mass 
available is uniquely determined
by the binding energy at the innermost stable orbit, and this fixes the
efficiency. A general relativistic treatment \cite{nt73} reveals that
$\epsilon$ increases from 0.06 (non-spinning black holes) to 0.42
(maximally rotating Kerr black holes).  
For radiatively efficient discs, the bolometric luminosity is 
characterized (and in general limited)
by the so-called {\it Eddington luminosity}: the luminosity
at which the radiative momentum flux from a spherically symmetric source is balanced by the 
gravitational force of the central object: 
$L_{\rm E}=\frac{4 \pi G m_{\rm p} M c}{\sigma_{\rm T}} \simeq 1.3 \times 10^{38} \frac{M}{M_{\odot}} 
\; {\rm erg \; s}^{-1}$,
where $m_{\rm p}$ is the proton mass and $\sigma_{\rm T}$ the Thompson 
scattering cross section.

When, on the other hand, radiative cooling is negligible, accretion is
adiabatic and $\epsilon \ll 1$. Then, unless
other mechanisms of energy redistribution are important, such as
convection or conduction, the
gravitational binding energy of the gas is converted into thermal
energy and merely advected with the flow (advection dominated
accretion flows or ADAF \cite{ich77,ny94}). However, any imbalance between
enthalpy, gravitational and kinetic energy can drastically modify the
nature of the flow, which can be dominated by convective motions 
(CDAF \cite{qg00a})
or result in substantial mass outflow \cite{bb99}.

I will begin this review with radiatively efficient discs, paying special
attention to the properties of the SS solution
(sec.~\ref{sec_ss}) and of its modification due to 
the effects of turbulent magnetic fields in the disc and the
generation of a magnetic corona (sec.~\ref{sec_cor}). After a brief discussion
of optically thin radiatively efficient solutions
(sec.~\ref{sec_sle}), I will turn to adiabatic flows, both optically
thin (sec.~\ref{sec_adaf}) and thick (sec.~\ref{sec_slim}),
concluding with a brief discussion of the role of
convection and outflows (sec.~\ref{sec_adios}). In sec.~\ref{sec_jet},
I sketch the possible relationships between different accretion modes
and MHD jet production mechanisms, before summarizing
(sec.~\ref{sec_map}). 
Most of the subjects touched upon here
are discussed thoroughly in more comprehensive reviews \cite{pl95,abr98,fkr02}.

\section{Radiatively efficient accretion discs}
\subsection{Optically thick flows: the standard solution}
\label{sec_ss}

The SS solution represents
an optically thick, geometrically thin, radiatively efficient
accretion flow.
In steady state, the disc structure  is determined by solving simultaneously four 
conservation equations (of vertical momentum, mass, angular momentum and energy).
If the equations of state for the pressure and opacity as functions of 
density and temperature are also given
together with a viscosity law, the full disc structure can be calculated exactly.
The flow is {\it assumed} to be geometrically thin, so that its pressure vertical scaleheight,
$H \equiv \frac{1}{2P}\frac{\partial P}{\partial z}$,
at a distance $R$, is much smaller than the radial coordinate, $H\ll R$.
In fact, since there is no {\it net} motion of the gas in the vertical direction, 
conservation of vertical momentum reduces to the equation of hydrostatic equilibrium,
from which we get  
$\frac{H}{R} \simeq \frac{c_{\rm s}}{v_{\rm K}}$,
namely the ratio of the disc height to the radius is approximately 
equal to the ratio of the local isothermal sound speed 
$c_{\rm s}=(P/\rho)^{1/2}$
to the local Keplerian velocity $v_{\rm K}=(GM/R)^{1/2}$.
The mass conservation equation for steady state accretion reads
$\dot M= - 2 \pi R \Sigma v_r$,
 where $\Sigma=\int^{+\infty}_{-\infty}\rho dz \simeq 2H \rho$ 
is the surface density and $v_r$ is the radial 
velocity of the spiraling matter.
Angular momentum conservation can be expressed by equaling the torque exerted by viscous stresses
to the net rate of change of angular momentum. This yields
$W_{R\phi}=\frac{\dot M \Omega}{2\pi \Sigma} J(R)$,
where $W_{R\phi}$ is the dominant ($R-\phi$) component of the stress
tensor, $\Omega=(GM/R^3)^{1/2}$  is the angular velocity and
the term $J(R)= \left[1-\left(\frac{R_{\rm in}}{R}\right)^{1/2}\right]$ 
comes from the (Newtonian) no-torque at the inner boundary
condition, and measures the rate at which the angular 
momentum is deposited onto the compact object.
From the energy conservation equation the surface emissivity can be calculated by taking the 
opposite of the divergence of the flux. This yields, for the heat production rate, 
\begin{equation}
Q_+ = \frac{3GM\dot M J(R)}{8\pi R^3},
\end{equation}
which is completely independent on the yet unspecified viscosity law.

Understanding the nature of the disc viscosity, and therefore understanding 
{\it how} can the disc accrete, 
has been (and partially still is) the central problem of accretion
disc theory.
Observationally, the accretion rates needed to
explain the luminosities we see are many orders of magnitude larger
than standard microscopic viscosities 
could provide. But if the disc is turbulent the effective viscosity due to
interacting eddies could easily be large enough \cite{bh98}.
The original SS solution relies on
a dimensional scaling for the turbulent viscosity coefficient,
$\nu=W_{R\phi}/\Omega=\alpha_{\rm v} c_{\rm s} H$, with the constant  
$\alpha_{\rm v} < 1$ (subsonic turbulence).

In geometrically thin accretion discs, the internally generated heat $Q_+$ is transported
vertically before being radiated at the surface. 
For optically thick
discs, photons are transported to the surface via diffusion, 
and the vertical photon flux is given by
$F_0=\frac{c}{\rho \bar \kappa}\frac{dP_{\rm rad}}{dz}$.
Therefore, the energy dissipation is locally balanced by radiation from the two disc surfaces, 
and we have 
$Q_- \equiv F_0 = Q_+$. Such a balance is
established over a thermal timescale, $t_{\rm th}\simeq 1/\alpha_{\rm
v}\Omega$. In thin discs, this is much smaller than the viscous time,
the time available to the accreting gas to radiate the energy released
by the viscous stresses: $t_{\rm v}\simeq (R/H)^2 J(R)/ \alpha_{\rm v} \Omega$.
Indeed, we are led the conclusion that SS 
discs are ``cool'' enough in that $kT\ll GMm_{\rm p}/R$, 
and therefore $H/R \ll 1$, as we had assumed. 

The final expressions for the physical quantities in the disc 
as functions of central mass, 
accretion rate, 
viscosity parameter and radius 
have simple algebraic expression when both total 
pressure and opacity are dominated by one term only.
As the free-free absorption is strongly suppressed at high temperatures,
electron scattering is the main source of opacity in the inner, hotter parts of the disc, 
while free-free absorption dominates in the outermost, colder parts.
Analogously, gas pressure dominates in the outer regions of the disc, while radiation pressure 
becomes predominant in the inner ones. Overall, three regions can be distinguished in a 
standard geometrically thin, optically thick accretion disc \cite{ss73}:
a) an inner part where electron scattering determines the opacity and radiation pressure
is larger than gas pressure; b) a middle region where electron scattering is still more important than free-free 
absorption, but gas pressure dominates over radiation pressure; c) an outer region, where thermal gas pressure dominates and free-free absorption is the main 
source of opacity.

\subsection{Beyond the standard model: Coupled magnetic disc--corona solutions
}
\label{sec_cor}

At present
magneto-rotational instability (MRI)\cite{bh98} is favoured as the
primary source of the turbulent viscosity needed to explain the
luminosities of accreting black holes. Our knowledge of the physics
of MHD turbulence in accretion discs can be used to build a
self-consistent model of MRI-driven, thin accretion disc--corona
systems in the following way.

Let us assume equipartition between kinetic turbulent energy and magnetic
field excited by the MRI (Alfv\'en equal to turbulent speed, $v_{\rm
A}=v_{\rm t}$). For almost
incompressible flows \cite{pbb02} the coefficient of turbulent viscosity
$\nu \simeq v_{\rm t}^2/3\Omega$, which implies $Q_+=\frac{3}{2}c_{\rm
  s} \nu \rho \Omega=c_{\rm s} P_{\rm mag}$.
Therefore, once the relationship between magnetic and disc pressure
(given either by gas or radiation) is established, the
accretion disc structure can be fully described.

To find such a relationship, we can assume \cite{me02} that the
magnetic field escapes from the thin disc via buoyancy, with a timescale $t_{\rm
  b}=H/2v_{\rm A}$. Then, a crucial point is that 
the growth rate of MRI is influenced by the ratio
of the gas to magnetic pressure \cite{bs01}: $\sigma=\Omega c_{\rm
g}/v_{\rm A}$, where $c_{\rm g}$ is the gas sound speed. The saturation
field can be found by noting that, asymptotically, $\sigma t_{\rm
b}=O(1)$; this automatically gives the desired scaling for the magnetic
pressure in MRI dominated turbulent flows:
\begin{equation}
\label{eq_visc}
P_{\rm mag}=\alpha_0 \sqrt{P_{\rm gas}P_{\rm tot}},
\end{equation}
where $\alpha_0=(\frac{P_{\rm tot}}{\beta^2 P_{\rm gas}})^{1/2}$ is a
constant, not necessarily smaller than unity.

The magnetic flux escaping in the vertical direction may dissipate a
substantial fraction of the gravitational binding energy of the
accreting gas {\it outside}  the optically thick disc, with obvious
deep implications for the spectrum of the emerging radiation.
The fraction $f$ of the total power dissipated in the low-density
environment above and below the disc (in the so-called {\it corona})
is determined by the ratio of the vertical Poynting flux ($F_{\rm
P}\simeq v_{\rm A}P_{\rm mag}$) to the local heating rate $Q_+$.
Under the assumption of  equipartition between turbulent and magnetic
energies, this translates into 
\begin{equation}
\label{eq_f}
f=\frac{v_{\rm A}}{c_{\rm s}}=\sqrt{\frac{2}{\beta}}
\simeq \sqrt{2 \alpha_0} \left(1+\frac{P_{\rm
      rad}(R)}{P_{\rm gas}(R)}\right)^{-1/4}.
\end{equation}
The ratio of the radiation to gas pressure can be calculated \cite{sz94} by simply
taking the SS solution and modifying it by reducing the locally
dissipated energy by the factor $(1-f)$. 
Then, for every value of the physical parameters 
$R$, $M$, $\dot M$ and $\alpha$, Eq.~(\ref{eq_f})
represents a nonlinear algebraic relation that determines $f$ and the
full solution of the accretion disc--corona system.
It turns out \cite{me02} that for $2\alpha_0<1$ such a solution is unique,
and has the following properties: $f$ tends to its maximum value,
$\sqrt{2\alpha_0}$,
when gas pressure dominates (low accretion rates), 
and decreases as the accretion rate
increases and radiation pressure
becomes more and more important\cite{mf02}. 
If instead $2\alpha_0>1$, there are no subsonic
(i.e. satisfying $v_{\rm t}<c_{\rm s}$) solutions for gas pressure
dominated regions, while in the radiation dominated part of the flow
two solutions are possible for every value of the accretion rate. The
first is a ``standard'' (unstable) one, with $f \ll 1$, while the second
appears due to the feedback effect of the closure relation (\ref{eq_f}).
This new solution, which is discussed 
in detail in \cite{me02}, has 
$f\rightarrow 1$ (corona dominated) 
as the accretion rate increases and is thermally and
viscously stable. It represent a new class of geometrically thin,
optically thick corona-dominated accretion discs, relevant for systems accreting at a
rate above the Eddington one.

\subsection{The SLE solution: radiatively efficient optically thin flows}
\label{sec_sle}
The first self consistent alternative to the standard solution for rotating accretion flows around black holes was 
discovered \cite{sle76} in an attempt to explain the hard spectrum of the classical black
hole candidate Cygnus X-1. This so-called SLE solution  is a {\it two temperature}
one, with the protons much hotter than the electrons $T_{\rm p} \gg
T_{\rm e}$. This is the case
if the energy released by viscous dissipation  was distributed equally among the carriers of mass, i. e.
mostly to the protons (which radiate very inefficiently) 
and only a fraction $m_{\rm e}/m_{\rm p}=1/1833$ to the electrons.  The hot proton component then 
dominates the pressure and keeps the disc geometrically thick. This in turn leads to low density
(which scales as $H^{-3}$ for given $\dot m$) and low Coulomb energy exchange between 
protons and electrons, that self-consistently implies $T_{\rm p} \gg T_{\rm e}$. The four main
equation of the disc structure are the same as in the standard model, 
but the heating--cooling balance is now described by two separate equations for the two species, where
the electrons mainly cool by up-scattering soft radiation coming from an outer cold disc or from 
dense cloudlets  embedded in the hot flow.

Although the model is able to reproduce reasonably well observed spectra of accreting black holes, 
it is thermally unstable \cite{pir78}: an increase in $T_{\rm p}$ would lead to disc expansion in
the vertical direction, further reduction of Coulomb cooling and
increase of proton temperature, leading to an instability.
This instability is cured if the effect of advection is taken into account in the energy equation, 
as will be briefly outlined in the following section.

\section{Adiabatic accretion flows}
\label{sec_ad}
\subsection{Optically thin solutions}
\label{sec_adaf}
It was first noted by Ichimaru \cite{ich77} that, 
in an optically thin, geometrically thick accretion disc such as the SLE one,
the density might be so low that the ions are unable to transfer
energy to the electrons in a timescale smaller than the viscous one. 
Then, part of the energy is advected with the proton flow and swallowed by the
central black hole. The energy equation then can be rewritten in a vertically integrated form:
$q_+-q_-=q_{\rm adv}\equiv \Sigma v_{\rm r} T \frac{\partial S}{\partial R}$,
where $S$ is the specific entropy of the gas.
This kind of solution was later named `ion supported torus'
\cite{ree82}, and it was also demonstrated
that it can be established only for accretion rates lower than a critical value
$\dot m < \dot m_{\rm crit} \sim  \alpha_{\rm v}^2$,
ensuring that the density is low enough, either because the accretion rate is low, or the viscosity 
high. The flow is optically thin and radiatively inefficient.
In recent years, much work has been devoted to the detailed study of
such flows, renamed
Advection Dominated Accretion Flows (ADAF) after Narayan and
Yi \cite{ny94} 
presented a full analytical self-similar 
1D solution to the problem.
This self-similar solution reveals the basic properties of an ADAF: 
a) the radial accretion time-scale is much shorter that that of a thin disc;
b) sub-Keplerian rotation occurs, due to large internal pressure
support; c) the flow is geometrically thick, in that the vertical
scaleheight $H\sim R$.
More uncertain, but crucial, issues for determining physical and
 observable properties of ADAFs are the assumptions
that ions and electrons interact only through Coulomb
collisions \cite{bc88} or  that 
only a tiny fraction of the turbulent energy is dissipated into the electrons \cite{bl97,qg99}.

\subsection{Optically thick solutions}
\label{sec_slim}

The last class of accretion modes comprises adiabatic solutions at accretion rates close to (or above)
the Eddington one. In these cases, the inflow 
timescale in the inner part of the disc becomes smaller than the time it takes for a 
photon to escape from the disc, and
the radiation produced is trapped by the flow and 
advected with it \cite{kat77,bm82}. The radius inside which radiation is trapped
moves outwards as the 
accretion rate increases\footnote{There is, though, the possibility
  that magnetized radiation pressure dominated discs are subject to
 ``photon bubble''  instabilities \cite{gam98} that lead to strong density
 inhomogeneities. As shown in \cite{beg01}, such discs may radiate
 well above the Eddington limit and remain geometrically thin without
 being disrupted.}. 
When it becomes of the order of the Eddington one, 
the disc thickness stays moderate and a vertically integrated  approximation may be retained
({\it slim discs}) \cite{abr88}. 
In the limit $\dot m \gg 1$, though, 
the behaviour of the disc and the possible relevance of strong outflows 
still remain open issues. Once again, the problem is inherently 2D, and
the simultaneous roles of convection, advection and outflows have to be assessed in order to 
model properly the expected SED. 
Unlike for the the optically thin ADAF/CDAF cases, the observable features of these
optically thick solutions have been investigated in only a handful of
cases so far \cite{szu96,mine00}, thus reducing
the general appreciation of their importance for interpreting observations.

\subsection{The role of convection and outflows in adiabatic flows}
\label{sec_adios}

The ADAF solution has received a great deal of interest in the last
decade because of its potential capability to explain the
observational data from our Galactic Center \cite{mf01,n02}.
On the other hand, geometrically thick adiabatic flows 
are ideal models to test global
numerical simulations of MHD turbulent accretion flows against, without
having to deal with the complications of radiative transfer 
needed to simulate thin, radiative efficient discs. 
Indeed, already when the 1D approximation of self-similar ADAF theory
is abandoned, and the full 2D nature 
of the problem is analysed, both from the theoretical point of view \cite{nia00} and from numerical
simulations \cite{ia00,hb02}, it is clear that radiatively inefficient flows are prone
 to strong convective instabilities and/or powerful outflows. 
 In general \cite{bb99,ia00}, convective flows are
more likely at low values of the viscosity parameter, while strong
outflows are generated for high values of $\alpha_{\rm v}$.
In the former case (purely convective flows), accretion is
effectively stifled, with little or no mass inflow or outflow: the
energy extracted in the inner part is convectively transported
outward. Such a redistribution of energy among fluid
 elements in the accreting gas alters the purely advective nature of
 the flow and modifies the radial profiles of physical quantities, as
 the density, with profound implication for the interpretation of the
 observed radiation.  
In the latter case, systematic outflows remove mass and energy from the
flow, with little accretion onto the black hole \cite{hbs01,hb02}. 
Despite the big efforts made by several groups, both on the theory
and on the simulation side, the relative importance of convection and outflow for adiabatic flows
is still matter of a vigorous debate \cite{bh02,nq02}, the controversy
being essentially over the capability of any hydrodynamical model
supplemented with $\alpha$-like viscosity prescriptions to capture
the basic physical properties of an inherently magneto-hydrodynamical system.
As usual, it appears likely
that such controversy will only be settled with the collection of more constraining
observational data.

\begin{figure}[ht]
\vbox to100mm{\vfil
\psfig{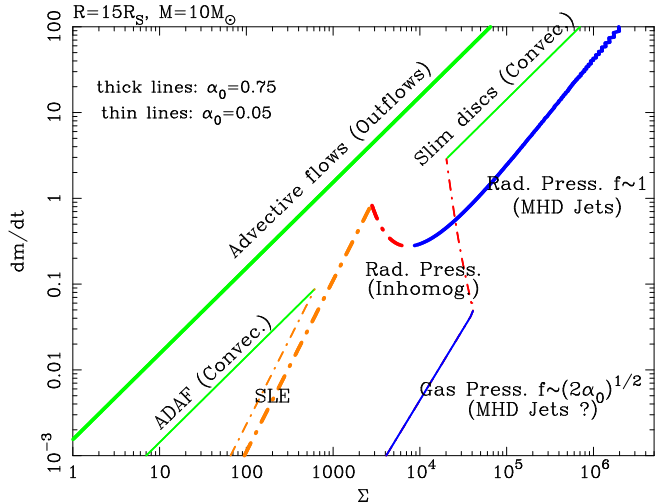}
\caption{Schematic representation of the different  
radiatively efficient and inefficient accretion flows, for 
$\alpha_0=0.05$ (thin lines) and $\alpha_0=0.75$ (thick lines), as functions of
accretion rate ($\dot m \equiv \epsilon _0 \dot M c^2/L_{\rm E}$, 
vertical axis) and surface density ($\Sigma$, horizontal axis). We
assumed $\epsilon_0=0.08$
(Newtonian disc, non-rotating hole), and fixed $M=10 M_{\odot}$,
$R=15R_{\rm S}$. 
Stable solutions are represented by solid lines, unstable ones by
dot-dashed line. Radiatively inefficient (stable) modes are in green,
optically thick, geometrically thin, radiatively efficient
(stable) solutions in blue, with indicated the dominant contribution
to the pressure and the value of the coronal fraction $f$. Unstable
(radiation pressure) thin disc solutions are in red, while SLE
solution are in orange. Also indicated in parenthesis are  additional
physical processed that are likely to be relevant for each kind of
accretion mode.} 
\label{fig_branches}
\vfil}
\end{figure}

\section{Accretion modes and jet engines}
\label{sec_jet}

It is a well established observational fact that accretion onto
rotating astrophysical object is often accompanied by the ejection of 
collimated outflows. Magnetic fields are likely to be a key element
for the understanding of such a phenomenon and a number of model of
magnetically-driven jets from rotating accretion flows have been
devised (for recent reviews see \cite{mku01,cb01}). Almost all of them link the power
channeled into the jet to the intensity of the poloidal component of
the magnetic field in the inner regions of the accretion
flow \cite{bz77,lop99}: $L_{\rm jet}\simeq \left(\frac{B_{\rm P}}{4\pi}\right)^2 A \Omega R$,
where $A$ represent the area of the acceleration region and $\Omega R$
is the rotational velocity of the field lines. 
Thus, the
kinetic output from a magnetized, rotating accretion flow depends on
two main factors: the scaleheight of the magnetically dominated
structure (because $B_{\rm p}/B_{\phi} \sim H/R$) 
and the maximal rotational velocity attainable.
The former should be large in all 
radiatively inefficient modes (in particular the more outflow-dominated 
ones at high viscosity parameter) 
and in those dominated by powerful magnetic coronae
(at low/high accretion rates for low/high viscosities, respectively). 
The latter depends on the angular momentum of
the central hole (high spin favors the generation of powerful
jets) and on the inner boundary condition for the
accretion flow, and can be tackled only by a full general relativistic
treatment of the inner disc. Different combinations of the two
factors may cause the variety of radio properties of AGN \cite{mei01}.

\section{Summary: a revised accretion map}
\label{sec_map}

Some of the properties of the different solutions can be summarized by the diagram
that places them in the  surface density ($\Sigma$)--accretion rate  ($\dot m$) plane 
(at fixed distance from the central source), 
as shown in Fig.~\ref{fig_branches}. For low values of the viscosity parameter $\alpha_0$,
the solutions split into two separate branches: the optically thin and the optically thick 
one. Optically thin solutions exist only at low accretion rates: they are the ADAF (thermally stable) and SLE (Shapiro-Lightman-Eardley, thermally unstable). Also 
at low $\dot m$, a geometrically thin, gas pressure dominated solution
exists, with an optically thin corona whose relative power is $f\sim \sqrt{2\alpha_0}$; 
this solution becomes unstable (and loses its corona) 
at the value of the accretion rate for which radiation pressure becomes larger than gas 
pressure  at the given distance from the black hole.
At even higher accretion rates only a radiatively inefficient, optically thick solution
exists (`slim disc'), which is thermally stable.

The topology of the diagram changes for high viscosity parameter: the advective, 
radiatively inefficient solutions (ADAF, slim disc) becomes a single branch, while 
radiatively efficient solutions form a second one. Of those, only corona-dominated, 
for accretion rates above a critical value, are stable. 

Fig.~\ref{fig_branches} also shows in
parenthesis the additional processes, beside those taken into account 
by the
model, likely to be relevant for each kind of
accretion mode. They are meant to be indicative of the open
theoretical issues in the field, and will likely indicate the
direction of the research in the immediate future.


\end{document}